\documentclass{svproc}

\usepackage{url}
\usepackage{verbatim}							
\usepackage{graphicx}
\usepackage{floatrow}
\usepackage{floatrow}
\usepackage[cmex10]{amsmath}
\usepackage[ruled,vlined]{algorithm2e}			

\begin{document}
\mainmatter             


\title{Morphogenesis of complex networks: a reaction diffusion framework for spatial graphs}

\titlerunning{Morphogenesis of complex networks}  

\author{ Michele Tirico\inst{1} \and Stefan Balev\inst{1} \and Antoine Dutot\inst{1} \and Damien Olivier\inst{1} }

\authorrunning{Michele Tirico et al.}

\tocauthor{Michele Tirico, Stefan Balev, Antoine Dutot and Damien Olivier}

\institute{Normandy Univ, UNIHAVRE, LITIS, 76600 Le Havre, France,
	\email{michele.tirico@univ-lehavre.fr} }

\maketitle              

\begin{abstract}
A large variety of real systems are composed by entities in relationships which can be represented by networks. 
In many of these systems, elements are embedded in the space and location information impacts properties and evolution. 
Local interactions between elements generate different kinds of equilibrium and often indicate a self-organized behaviour. 
In this paper we are interested in essential mechanisms behind morphogenesis of spatial networks such as street networks. We propose a multi-layer model, where a reaction-diffusion mechanism governs the growth of spatial networks. We study its evolution with some metrics.
\end{abstract}

\section{Spatial networks: properties and models}
Network theory is a relevant methodology to represent, describe and analyze complex systems. It provides a powerful representation for many social, biological and technological applications \cite{latora_complex_2017,boccaletti_complex_2006,batty_cities_2007}. 
In some cases, geographical localization plays a crucial role to describe essential properties.
Compared to other complex networks, spatial networks often show higher robustness, less peaked and more compact degree distribution \cite{barthelemy_spatial_2011}.
Spatial networks are widely used to represent venation pattern of leaves \cite{katifori_quantifying_2012}, insect colonies \cite{perna_characterization_2011} and street networks\cite{viana_simplicity_2013}.
In urban studies, centrality metrics are useful to identify loops \cite{lion_central_2017}, patterns \cite{jiang_topological_2004},  accessibility \cite{porta_network_2006} and its historical evolution is largely studied with topological indicators \cite{masucci_random_2009}. The last ones can characterize some static properties but it is difficult to catch the dynamics with them. Our aim is to explore a simple model able to capture the development of a spatial graph under constraints.

The location of elements and the spatial dependencies should be taken carefully into account in order to identify an appropriate methodology to generate spatial networks.
Accordingly, investigating essential evolution mechanisms of spatial networks is therefore interesting both for purely graph theoretical research and for a large number of real case studies \cite{barthelemy_morphogenesis_2018}.
In last decades, following the concept that the growth of spatial networks is subject to certain optimization processes \cite{gastner_spatial_2006}, various authors proposed models that yield to minimize metrics \cite{guillier_optimization_2017}, reproduce emergent characteristics \cite{barthelemy_optimal_2006} or find compromise between antagonist properties \cite{brede_coordinated_2010}. 
Using a global optimization process, interesting works are developed in order to replicate urban growth.
For example, geometrical optimization of global elements of urban network used in \cite{courtat_mathematics_2011} reproduces a coherent and realistic growth of cities.

Although global approaches are largely useful to model complex system evolution, in many applications of spatial networks, local dynamics are predominant and must be considered to produce emergence of global properties.
The formation of stable structures is often a consequence of elementary local interactions. They are dependent on spatial evolution of elements in a small neighbourhood \cite{nicolaides_self-organization_2016}. 
A class of structures in dynamic equilibrium emerges as a result of local interactions, not through external and global control. 
Through a local-driven interaction approach, the network grows spatially and temporally in a coherent way. 
In many applications, spatial and temporal coherence is strongly required: for instance, in urban studies, a typical street network is growing as a single connected component \cite{achibet_model_2014}. 

The generation of disconnected sub-graphs is often not an appropriate representation of growth for some applications. For example, vascular systems or insect gallery networks must always be connected. Moreover, in venation leaves and street growth pattern formation, the network evolution thanks to local interactions produces networks which often respect the planarity condition. Roughly, a planar graph is a spatial graph that can be drawn in a plane without edge crossing \cite{diestel_graph_2010}.  
Growing planar graph models with focus on local interactions are proposed in urban studies: Barthelemy et al. \cite{barthelemy_modeling_2008} describe a model of street network growth with local optimization of shortest connection road and an handle formation of loops. In Rui et al. \cite{rui_exploring_2013} model, competition of nodes added step-by-step produces a stable self-regulated street network.

The study of spatial network evolution is a fundamental topic with many applications. The purpose of this paper is to formalize a minimal model, using different kind of local interactions and where evolution is self-regulated, in order to investigate essential mechanisms of morphogenesis and capture the complexity of the system. In section 2 we expose the framework of our model, its essential elements and interaction mechanisms. In section 3 we show and discuss early results through some metrics. 

\section{Model description}
Morphogenesis is the process of development of the forms of a living organism and by extension of some human organizations like cities. In 1952, Alan Turing has proposed a mathematical model \cite{turing_chemical_1952} of spatial structure morphogenesis. In this model there are chemical compounds which react together and diffuse. The result is the production of evolving patterns in a continuous medium. Two molecules (\verb|A| and \verb|B|) called morphogens interact: \verb|A| catalyzes its own production but also the production of \verb|B|. In the same time \verb|B| inhibits the production of \verb|A|. \verb|B| diffuses faster than \verb|A|.

\floatsetup[figure]{capposition=beside,capbesideposition={bottom,right}}
\begin{figure} [t]
	\centering
	\includegraphics [trim = 0cm 0cm 0cm 0cm , scale=.8 ]	{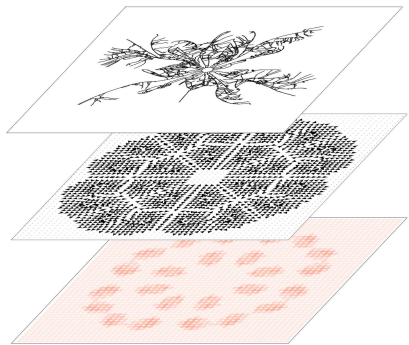}
	\caption{ 
		The model is composed of three interdependent layers: a regular grid governed by a reaction-diffusion mechanism (bottom), a vector field applied to a set of moving seeds (center) and a growing spatial network (top).
	}	
	\label{fig_layers3d}
\end{figure}

Due to the auto-catalytic reaction and different diffusion rates of morphogens, this essential framework is able to produce a variety of spatio-temporal patterns. 
Through this minimal framework, many studies had been proposed to explain various spatial phenomena in continuous media, fish skin, sea shells and bacteria aggregation \cite{kondo_reaction-diffusion_2010}. 

Morphogens, in a real case study such as the urban growth, should represent a set of spatial properties, for example natural constraints, human behaviours, socio-economical effects.
The different concentration of these evolving properties impacts the evolution of the street network through constraints or attracting spots located in the space.
A set of forces, modelled as a vector field, governs the morphogenesis of the network. 
Therefore in this paper, we propose a growing spatial network generated by a discrete self-regulated pattern formation model. The model is composed by three interdependent layers:

\begin{itemize}
	\item a reaction-diffusion system applied to a square lattice, 
	\item a dynamic vector field applied to a set of moving seeds,
	\item and a growing spatial graph (fig. \ref{fig_layers3d}).
\end{itemize}

The first layer of our model dynamically governs the morphogenesis of the growing graph through spatial evolution of two morphogens in two dimensional space. 
Among different formalizations of the Turing concept, the Gray-Scott model, thanks to the introduction of a third mechanism (to increase or decrease the morphogen concentration), produces a wealth of distinct patterns \cite{gray_autocatalytic_1983}. 
The first layer of our model is inspired by latter studies. 
The continuous system of equations of Gray-Scott model, represents the evolution of morphogen concentration:

\begin{equation}
\begin{cases} 
\label{eqRdmContinous}
\frac{ \partial a }{ \partial t } = D_a \nabla ^2 a - ab^2 + f ( 1 - a ) \\	
\frac{ \partial b }{ \partial t } = D_b \nabla ^2 b + ab^2 - (f + k ) b 
\end{cases}	
\end{equation} 

where $a=a(X,t)$ and $b=b(X,t)$ indicate the concentration of two morphogens ($a$ and $b$) at time $t$ at position $X = (x,y)$ and the diffusion rate is computed through Laplacian terms  $\nabla ^2 a$ and $\nabla ^2 b$, incremented by  $D_a$ and $D_b$ for $a$ and $b$ respectively. Free parameters $f, k$ represent diminished and increment terms, both proportional to concentration of morphogens.

Based on the microscopic master equation (\ref{eqRdmContinous}) we use a discrete equation system to reproduce the evolution of each morphogens. 
The reaction-diffusion layer $C = [a(X,t), b(X,t)]$ in our model is a lattice of square cells, in which morphogen concentration in each cell is calculated as follow:

\begin{equation}
\begin{cases}
a(X\!,\!t\!+\!1)\! =\! a(X\!,\!t)\! +\! D_a \sum_{Y\!\in\! N(X)}\!{(a(X\!,\!t)\! -\! a(Y\!,\!t))}\! -\! a(X\!,\!t) b(X\!,\!t)^2\! +\! f(1\!-\!a(X\!,\!t)) 
\\
b(X\!,\!t\!+\!1)\! =\! b(X\!,\!t)\! +\! D_b \sum_{Y\!\in\! N(X)}\!{(b(X\!,\!t)\! -\! b(Y\!,\!t))}\! +\! a(X\!,\!t) b(X\!,\!t)^2\! -\! (f\!+\!k)b(X\!,\!t)
\end{cases}	
\label{eq:reactiondiffusion}
\end{equation} 

where $N(X)$ is the Moore neighbourhood of the cell at position $X$ \cite{kari_theory_2005}.
The diffusion rate is obtained through the Fick's law, to ensure a local mass equilibrium \cite{bird_theory_1956}. 

The different concentration of morphogens can be seen as a set of attraction or repulsion spots for a set of moving elements, in the following called seeds. 
A force depending on the distance and the morphogen concentration moves the seeds.
Accordingly, in a discrete representation, we compute a vector field $B = ( \vec{b}, t ) $ for the morphogen $b(X,t)$. At each cell on layer $C$ is associated the vector:

\begin{equation}
\vec{b} ( X, t) = \sum_{Y\in N(X)}  { g \frac{b(X,t) b(Y,t)}{d(XY)^\theta} \hat{XY} }
\label{eq:vectorfield}
\end{equation}

where $	{d(XY)} $ is the euclidean distance between centres of cells located in $X$ and $Y$, $\hat{XY}$ the relative unit vector,  $g$ is a gravitational constant and $\theta$ is a distance exponent. 
The second layer is completed by an evolving set of moving seeds $S = [s_1(X_1,t),.., s_n(X_n,t)]$. During the simulation, the vector field is applied to the moving seed set and each element follows the evolution of the gradient of morphogen concentration.  
To each seed we apply a vector $\vec{r}$ (the sum of corresponding nearest vectors) and the evolution of positions of the seeds represents the latest layer of the model.
More precisely, we generate a growing spatial graph $G=(V(X,t) ,E(t))$, a mathematical entity composed of an evolving set $V$ of vertices embedded in a two dimensional space, and a set $E$ of undirected edges \cite{holme_modern_2015}. 
Using rules exposed in the following, the graph grows spatially and temporally in a coherent way, without edges crossing.

\subsection{Spatial interactions: the morphogenesis of the network\label{subsec:spatialinteractions}}

\floatsetup[figure]{capposition=bottom}
\begin{figure} [t] 		
	\centering
	\includegraphics [trim = 0cm 0cm 0cm 0cm , scale=.7 ]	{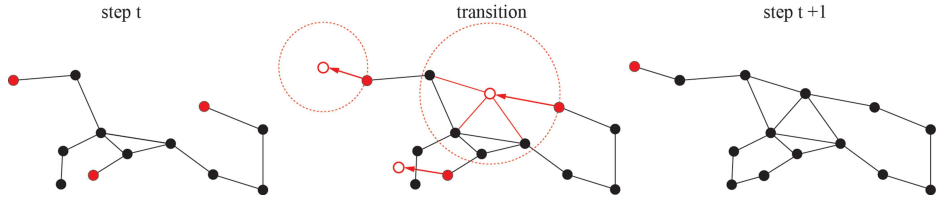}
	\caption{ 
		Interactions between seeds and the graph. In the step $t$ (left), due to the corresponding resultant vector, seeds move and link to the previous position and all nodes in the radius; if a seed crosses an edge, it connects to the nearest node (center). At step $t+1$, the result of these interactions is a connected graph without crossing edges (right).     
	}	
	\label{fig_vectors}
\end{figure}

\floatsetup[figure]{capposition=beside,capbesideposition={bottom,right}}
\begin{figure} 
	\centering
	\includegraphics [trim = 0cm 0cm 0cm 0cm , scale=0.75 ]	{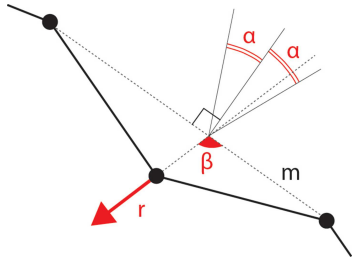}
	\caption{ 
		The seed creation. At each time step, whether the difference between directions of vector r (from all nodes with degree 2) and m (the sum of corresponding nearest vectors) is close to right angle, we add a new seed.
	}	
	\label{fig_angleSeed}	
\end{figure}

The dynamic vector field affects the position of moving seeds, which generate the graph. However, the morphogenesis of the network is not only governed by the reaction-diffusion mechanism. During the simulation, the network grows and its configuration impacts its own morphogenesis.

We start with a small connected graph with a seed in each node. 
At each seed corresponds a unique vertex in the growing graph. At each time step, we apply the corresponding vector $\vec{r}$ to each moving seed and calculate its future position $X_f$ as $X_t$ moved by $\vec{r}$. 
In this way we take into account the evolution of morphogen concentration near the seeds. 
Concerning the seed movement, several cases can occur; given $u$ a node at position $X_t$ and $v$ a future node at position $X_f$ :

\begin{itemize}
	\item If the potential edge between $u$ and $v$ crosses another edge, we do not create node $v$, we connect $u$ to the nearest node, and remove the seed.
	\item Otherwise we create node $v$ and the edges between $u$ and $v$. \begin{itemize}
		\item If there are no other nodes in a radius $\|\vec{r}\|$ around $v$ we keep the seed,
		\item otherwise we connect $v$ with each other node in radius $\|\vec{r}\|$ and remove the seed. (fig. \ref{fig_vectors})
	\end{itemize}
\end{itemize}

Seeds are not only removed, but also created during the simulation. We create a seed at a node $u$ of degree two if the corresponding vector $\vec{r}$ is almost perpendicular to the line passing between the two neighbours of $u$ (fig. \ref{fig_angleSeed}). The parameter $\alpha$ is the maximal allowed deviation from the right angle.
The aim of the last mechanism is to consider the evolution during the simulation of the reaction-diffusion layer, in order to handle the generation of new seeds. The vector field evolve during the simulation and the corresponding nearest vector of the vertex in the growing graph change intensity and direction: each vertex in the graph could be a potential source of seeds as consequence of combination of the vector field and the network geometry.

Thanks this methodology we obtain graphs which grow coherently, embedded in two-dimensional space and where edges do not cross each other. 
The simulation starts with the creation of three layers and ends when the moving seed set becomes empty or after a predefined number of steps.
At each time step, we update the reaction-diffusion layer and compute the corresponding vector field.
The graph evolves thanks to the addition of new seeds and the movement of seeds, respecting local rules of interaction.
The pseudo-code of simulation is shown in Algorithm \ref{algoSim}. 

\begin{algorithm} [H]
	set $C$ a reaction-diffusion layer \\
	set $B$ a vector field computed from $C$  (see eq. \ref{eq:vectorfield}) \\  
	set $S = \{s_1,..., s_n\}$ a moving seed set \\ 
	set $G = (V(X), E)$ an initial graph \\
	\While() {$ S \neq \emptyset $ }{  
		update the reaction-diffusion layer $C$~(See eq. \ref{eq:reactiondiffusion}) \\
		update the vector field $B$~(See eq. \ref{eq:vectorfield}) \\		
		\ForEach (\tcc*[f] {Add seeds, see fig. \ref{fig_vectors}}){vertex $v \in V(X)$ with degree $2$  } { 
			$\vec{r} \leftarrow$ sum of nearest vectors \\
			$\vec{m} \leftarrow$ vector between the two neighbours of $v$ \\
			$\beta \leftarrow$ angle between $\vec{r}$ and $\vec{m}$ \\
			\If  {$|\beta\%2\pi| \leq \pi/2 \pm \alpha$ } {
				$s \leftarrow$ a new seed at $v$ \\
				$S \leftarrow S + s$
			}
		}				
		\ForEach(){ $ s \in   S$  }{
			$v \leftarrow$ vertex where the seed $s$ is located\\
			$X_s \leftarrow$ current position of seed $s$ \\
			$\vec{r} \leftarrow$ as the sum of nearest vectors \\
			$X_f \leftarrow$ $X_s$ moved by $\vec{r}$ \tcc*[f]{future position of seed $s$} \\	
			\eIf(\tcc*[f] {Check edge crossing}){line $\overline{X_s X_f}$ intersects edge segments}{
				get nearest vertex $u  \in V $ from $s$ \\
				$E \leftarrow E + (u,v)$ \tcc*[f]{Add edge}\\
				$S \leftarrow S - s $ \tcc*[f]{Remove seed}\\
			}{
				$u \leftarrow$ new vertex at $X_f$\\
				$V(X) \leftarrow V(X) + u$  \\
				$E \leftarrow E + (u, v)$ \tcc*[f]{Connect new vertex to previous one}\\
				Move seed $s$ to $X_f$  \\
				$N(X) \leftarrow$ set of vertices within a distance $\left\lVert\vec{r}\right\rVert$ from $X_f$ \\		
				\If{$N(X) \neq \emptyset$}{		
					\ForEach(\tcc*[f] {Connect vertex to nearest vertices}){vertex $v(x_t) \in N(X)$ }{
						\If{line $\overline{X_t X_f}$ does not intersect edge $e \in E$}{		
							$E \leftarrow E + (v(x_t), v)$ \\
						}
					}
					$S \leftarrow S - s$ \\
				}	
			}
		}
	}
	\caption{Simulation	\label{algoSim}}
\end{algorithm}

\floatsetup[figure]{capposition=bottom}
\begin{figure}  [t]
	\centering
	\includegraphics  [trim = 0cm 0cm 0cm 0cm , scale=1.1 ]	{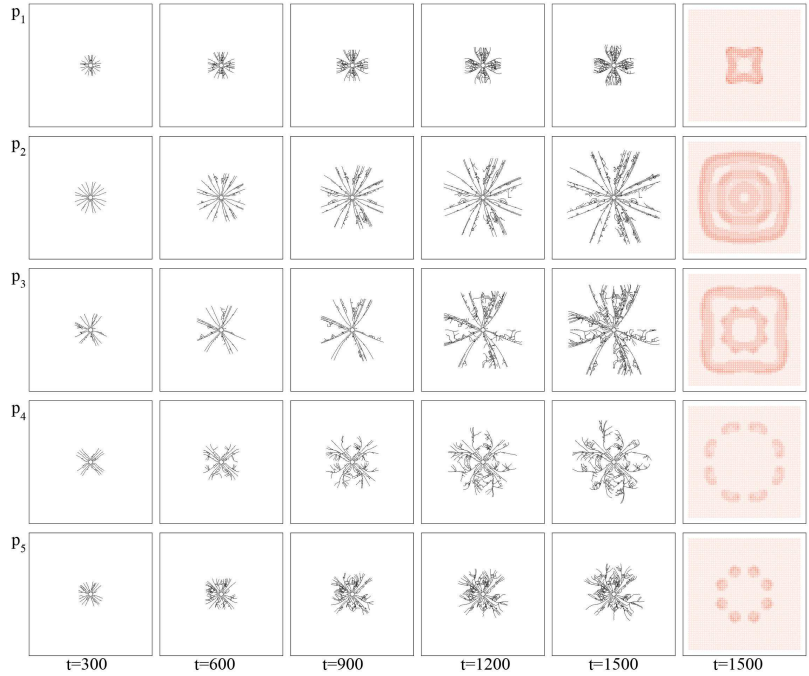}
	\caption{ 
		Simulations of five growing graphs and relative pattern, with $\alpha = 1.0^\circ $.
		Each row shows the corresponding pattern at final time step $t = 1500$ (last column) and the corresponding growing graph every 300 time steps.
	}	
	\label{fig_matrixRdmStep}
\end{figure} 

\section{Simulation and results}
The code has been developed in Java and we use the library GraphStream \cite{pigne_graphstream:_2008} to represent our networks and for some differents measures on them.
The reaction diffusion layer consists of a lattice of 50x50 unitary cells.
The start concentration of morphogens is trivial state $a=1.0$ and $b=0.0$, with a small localized perturbing pulse ($b=1.0$), a given number of moving seeds and associated nodes in the middle of the space.
Diffusion parameters are $D_a = 0.1$, $D_b = 0.2$, feed and kill parameter set ($f,k$) is been chosen in order to obtained some classical patterns:  $p_1$  (0.055 , 0.062), $p_2$  (0.039, 0.058), $p_3$  (0.029, 0.057), $p_4$ (0.014, 0.054), $p_5$ (0.025, 0.060) \cite{adamatzky_generative_2018}.
At the center of each cell in the reaction diffusion layer we compute the vector $\vec{b}$ following the eq. \ref{eq:vectorfield},  where $\theta= 2$ and $g=1$. 
For each seed, the corresponding vector $\vec{r}$ is the sum of the four nearest vectors. 
For each pattern $p$, we varied the free parameter $\alpha =[0.1^\circ, 0.2^\circ, ... 1.0^\circ]$ and we obtained fifty growing networks 
\footnote{Videos of 3 layers are computed for $\alpha=1.0^\circ$ until step $t=2500$, saved every 25th iterations; playback is 3 frames per seconds. \\
	Pattern $p_1$: https://youtu.be/2izGpD2XU0w \\
	Pattern $p_2$: https://youtu.be/IwG3oSewSpI \\
	Pattern $p_3$: https://youtu.be/ceQVYPadENY \\
	Pattern $p_4$: https://youtu.be/LMn6vv9dy7Q \\
	Pattern $p_5$: https://youtu.be/vMiAC5rZpzs
}.

\begin{figure} 
	\centering
	\includegraphics [width=1\linewidth]
	{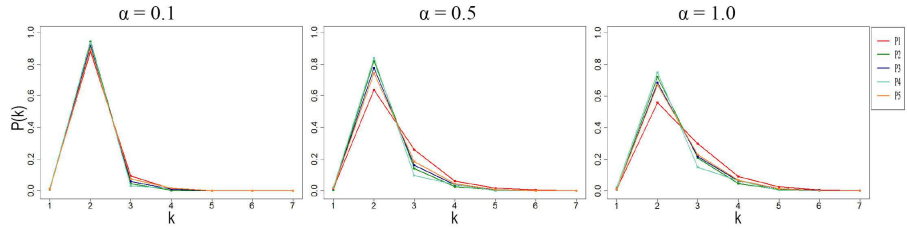}
	\caption{ 
		The degree distribution of five growing graphs calculated for all patterns and for three values of $\alpha =[0.1^\circ, 0.5^\circ, 1.0^\circ]$.			
	}	
	\label{fig_degreeDistribution}
\end{figure} 

\begin{figure} [t] 
	\centering
	\includegraphics [width=1\linewidth]	{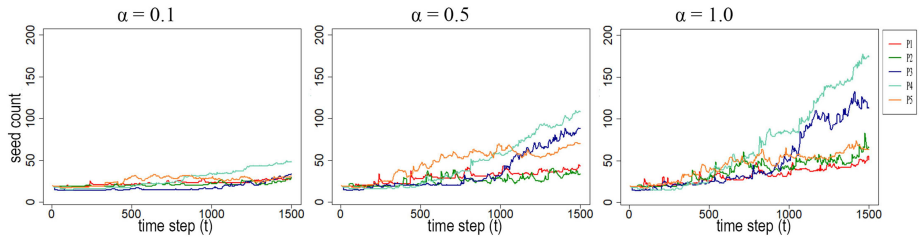} 
	\caption{ 
		The evolution of number of moving seeds, calculated for all patterns and for three values of $\alpha =[0.1^\circ, 0.5^\circ, 1.0^\circ]$.			
	}	
	\label{fig_degree1}
\end{figure}

\begin{figure} [t]
	\centering
	\includegraphics [width=1\linewidth]	{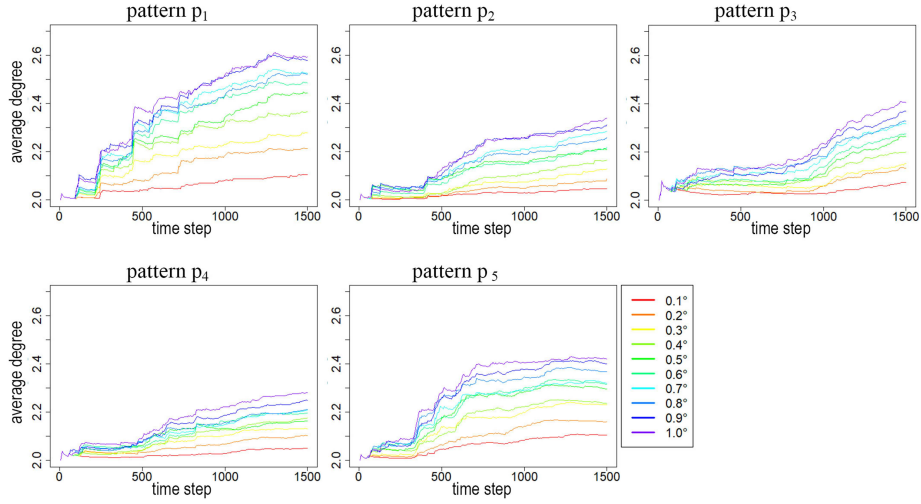}
	\caption{
		The evolution of the average degree calculated for all patterns and for all values of $\alpha =[0.1^\circ, 0.2^\circ, ... 1.0^\circ]$.
	}	
	\label{fig_averageDegree}
\end{figure}

\begin{figure} [t]
	\centering
	\includegraphics [width=1\linewidth]	{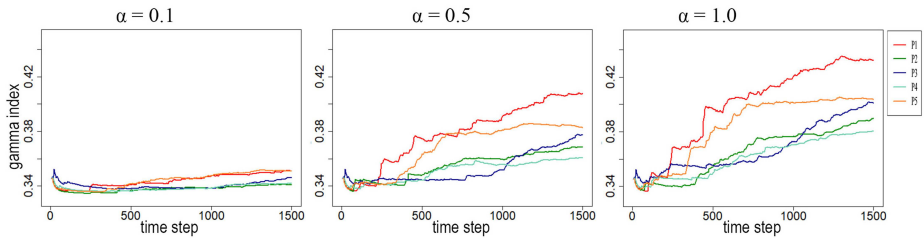}
	\caption{ 
		The evolution of gamma index computed for all patterns and for three values of $\alpha =[0.1^\circ, 0.5^\circ, 1.0^\circ]$.
	}	
	\label{fig_gammaIndex}
\end{figure}

\begin{figure} [t] 
	\centering
	\includegraphics [width=1\linewidth]	{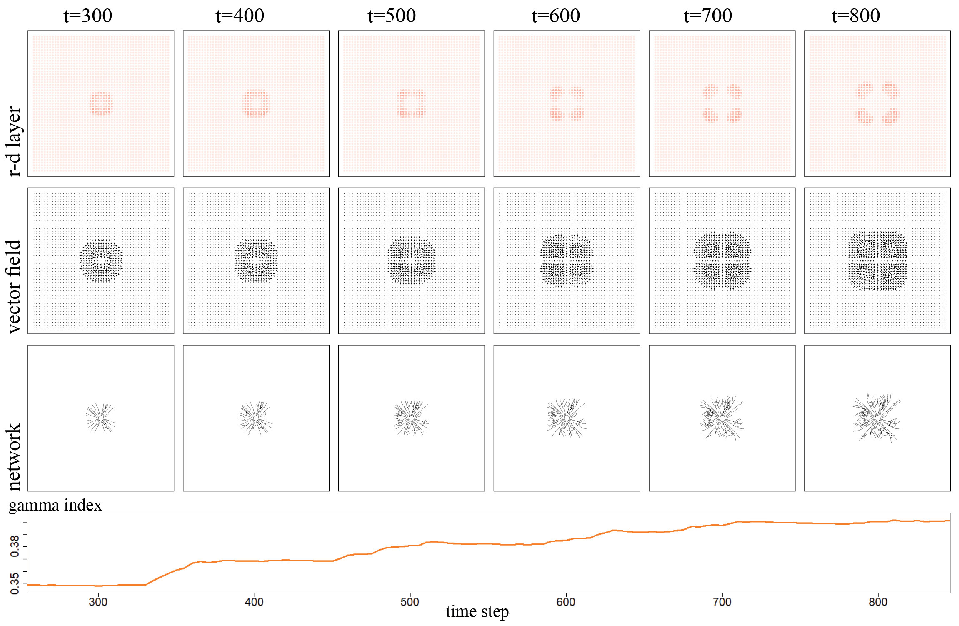}
	\caption{ 
		Snapshots (from step $t=300$ to $t=800$, computed each 100 steps) of the three layers (the first three rows) for the pattern $p_5$ and the corresponding evolution of $\gamma$ index (the fourth row).
	}	
	\label{fig_mitosis}
\end{figure}

In this section, we study some preliminary properties of all growing graphs generated through the model with few typical topological metrics largely applied in spatial graph analysis \cite{buhl_topological_2006}.
Our interest focus on three aspects: investigate patterns effects on growing networks, evaluate the evolution of graph topology and the impact of the $\alpha$ parameter.   

In fig. \ref{fig_matrixRdmStep} we show growing networks every $300$ steps, corresponding to reaction-diffusion patterns $p$ with the relative concentration of morphogen $b$ at step $t = 1500$; each growing graph is obtained with $\alpha = 1.0^\circ$. 
We observe that growing networks is strongly dependent to related reaction-diffusion patterns. 

The probability of finding vertices with degree $k$ is $ P(k) = v(k)/v $ (where $v(k)$ is the number of vertices with degree $k$ and $v$ is the total number of vertices). 
In most planar graph applications, the degree of the vertices is comprised between 1 and 7. In our case study, degree distribution shows an hierarchical law behaviour and follows a fast decay from $k=2$ to $k=6$, typically observed in street, leaf and ant gallery networks (fig. \ref{fig_degreeDistribution}).
In addition, the high probability to generate a new seed, suggested by the frequency $P(2)$, is not actually correlated to real generation of new seeds.
During the simulation, only few vectors related to vertices with $k=2$ permits to generate a new seed.
Fig. \ref{fig_degree1} depicts the  evolution of the number of seeds for three values of  $\alpha =[0.1^\circ, 0.5^\circ, 1.0^\circ]$.   		

In fig. \ref{fig_averageDegree} we show the evolution of average degree $ \left\langle k \right\rangle = 2e / v $ for all patterns $p$ for all value of  $\alpha$ parameter. 
As expected, parameter $\alpha$ and pattern $p$ play crucial role on the evolution of the average degree $\left\langle k \right\rangle $. 
We observe a similar behaviour for all patterns.
The $\alpha$ parameter impacts the connectivity at global scale and a sudden variation is detected at the same time step for each graph.
Results suggest that the evolution of morphogens concentration governs the growth of the network and the parameter $\alpha$ amplifies graph average degree.

In order to characterize the evolution of graph density, we calculate the gamma index. 
It is defined as $ \gamma = e / e_{MAX}$ and compares the actual number of edges $e$ to theoretical maximum number of edges  $e_{MAX} = 3(v - 2) $ (in the case of undirected planar graph) \cite{xie_geographical_2007}. 
In fig. \ref{fig_gammaIndex} we observe ample fluctuations during the simulation.
Like for average degree $\left\langle k \right\rangle $ evolution, we observe non monotone curves for all patterns, as a consequence of an unstable process.
Accordingly to a sudden change of morphogen concentration $b$, graphs grows discontinuously.
For instance, observing spatial evolution of concentration $b$ in pattern $p_5$, a mitosis process increase number of unstable spots (fig.~\ref{fig_mitosis}). 
In the reaction diffusion layer, this process causes the interaction showed in fig. \ref{fig_angleSeed} and increases the number of seeds. The network suddenly grows and the $\gamma$ index shows the rise of the graph density.
This clearly demonstrates that growing network mechanisms proposed in this paper consider at all times the evolution of morphogens concentration and the spatial properties of the network.  

\section{Conclusion}
In this paper we have exposed a methodology able to generate graphs embedded in two-dimensional space. 
A reaction-diffusion system, applied to a regular grid, governs a dynamic vector field, which impacts a set of moving seeds.
Growing graphs respect the planarity condition, connecting the new vertices to elements in an evolving neighbourhood.
The parameter $\alpha$ affects network growth, through local interactions between graph elements and the vector field.

Thanks to simple local connection rules, which govern spatial interactions between moving seeds and the growing network, we have developed a parsimonious model, where spatial and temporal coherence is ensured during the simulation and generate only one connected graph.
Combined with the planarity condition our model can be used in a wide range of applications.
The evolution of technical spatial networks such as street networks should be an interesting application for a framework governed by elementary local interactions.

Although these preliminary and not exhaustive evidences, several questions remain unanswered and required an in-depth investigation.
Our model allows to define different levels of abstraction.
For instance, in urban studies, it may represent sprawl of a single city at microscopic scale or the densification of connections between cites at macroscopic scale.
Our minimal approach integrates few essential physical processes and permits to investigate elementary mechanisms of spatial graph generation.

These early results should be integrated in the next works by a systematic evaluation of spatial effects of graph growth to evolving concentration of morphogens, in order to take into account the feedback of the graph to the reaction-diffusion layer. In a complex system, like a growing city, the network plays a crucial role in the spatial evolution of forces acting on the development of the city.

\paragraph{\textbf{Acknowledgements}}
This work is supported by the project “AMED” co-funded by ERDF and the region Normandy.

%
\bibliographystyle{spmpsci}
\bibliography{conf_cna_2019}

\end{document}